\documentclass[]{ptephy_v1}

\preprintnumber{XXXX-XXXX} 
\usepackage{hyperref}
\usepackage{orcidlink}
\usepackage{amsmath}
\usepackage{graphics}
\usepackage{url}
\newcommand{\python}{{\sc python}}
\usepackage{graphicx}
\usepackage{dcolumn}
\usepackage{bm}
\usepackage[utf8]{inputenc}
\usepackage[T1]{fontenc}
\usepackage{bm}
\usepackage{xcolor}
\usepackage{enumerate}
\usepackage{tabularx}
\usepackage{upgreek}
\usepackage{siunitx}
\usepackage[para]{threeparttable}
\def\titlepagefootline{} 

\hypersetup{
    colorlinks=true,
    linkcolor=blue,
    citecolor=blue,
    filecolor=blue,      
    urlcolor=blue
    }

\begin{document}

\title{Demonstration of Efficient Radon Removal by Silver-Zeolite in a Dark Matter Detector}


\author[1,*]{D.~Durnford, \orcidlink{0000-0002-6608-7650}} 
\author[1,*]{Y.~Deng, \orcidlink{0009-0009-3378-5338}}
\author[1]{C.~Garrah\footnote{now at Department of Physics, Engineering Physics, and Astronomy, Queen’s University, Kingston, Ontario, Canada}}
\author[1]{P.~B. O'Brien\footnote{now at Department of Earth and Atmospheric Sciences, University of Alberta, Edmonton, Alberta, Canada}}
\author[2]{P.~Gros}
\author[3]{M.~Gros}
\author[4]{J.~Busto}
\author[5]{S.~Kuznicki}
\author[1,*]{M.-C.~Piro, \orcidlink{0000-0003-3972-2708}}

\affil[1]{Department of Physics, University of Alberta, Edmonton, AB, T6G 2E1, Canada}
\affil[2]{Department of Physics, Engineering Physics, and Astronomy, Queen’s University, Kingston, Ontario K7L 3N6, Canada}
\affil[3]{IRFU, CEA, Université Paris-Saclay, F-91191 Gif-sur-Yvette, France}
\affil[4]{Aix-Marseille Université, CNRS/IN2P3, CPPM, Marseille, 13288, France}
\affil[5]{Department of Chemical and Materials Engineering, University of Alberta, Edmonton, AB T6G 2R3, Canada
\email{\textcolor{blue}{ddurnfor@ualberta.ca}}, \email{\textcolor{blue}{ydeng6@ualberta.ca}}, \email{\textcolor{blue}{mariecci@ualberta.ca}}
}

\begin{abstract}
We present the performance of an efficient radon trap using silver-zeolite Ag-ETS-10, measured with a spherical proportional counter filled with an argon/methane mixture. Our study compares the radon reduction capabilities of silver-zeolite and the widely used activated charcoal, both at room temperature. We demonstrate that silver-zeolite significantly outperforms activated charcoal by three orders of magnitude in radon capture. Given that radon is a major background contaminant in rare event searches, our findings highlight silver-zeolite as a highly promising adsorbent, offering compelling operational advantages for both current and future dark matter and neutrino physics experiments. Furthermore, this not only offers great promise for developing future radon reduction systems in underground laboratories, but also paves the way for innovative, multidisciplinary advancements with far-reaching implications in science, engineering, and environmental health.
\end{abstract}

\subjectindex{C43}

\maketitle

\section{Introduction}
Radon ($^{222}$Rn), a naturally occurring radioactive noble gas produced from the decay of $^{238}$U, poses significant challenges to low-background detection techniques in particle physics. As a major contaminant, radon compromises the sensitivity of experiments such as dark matter searches and neutrino physics studies \cite{ODwyer:2011, Pollmann:2012, GERDA:2013krg, Amaudruz:2012hr, Nakano:2017rsy, Miller:2017tpl,  Benato:2017kdf, Rupp:2017zcy, Street:2017bde, Quintana:2018, Monte:2018vjt, NEXT:2018zho, CRESST:2019oqe,  Lehnert:2018cpv, Zhang:2020qfs,LZ:2020fty, Borexino:2019wln, nEXO:2021ujk, DAMIC:2021crr, XENON:2021mrg,  Hussain:2022, Arthurs:2022, PandaX-4T:2021lbm, Ha:2022psk, SNO:2022qvw, Street:2023, CUPID-Mo:2023vle, LZ:2022ysc, LZ:2022lsv, Chott:2022lnc, Veeraraghavan:2023vxd,  MicroBooNE:2023sxs, Seo:2024rea, XENON:2024lbh, Cui:2024ltd, MillerChikowski:2024, XLZD:2024nsu, XLZD:2024pdv, Lahaie:2024jty}. $^{222}$Rn decays by emitting an alpha particle with an energy of 5.49 MeV and has a half-life of 3.82 days \cite{nndc_218}. Its decay chain includes two other alpha emitters, $^{218}$Po and $^{214}$Po, before finally producing $^{210}$Pb, which is also a significant background source \cite{XMASS:2017sid, Bunker:2020sxw, NEWS-G:2020fhm, Vivo-Vilches:2021lwr, DAMIC:2021crr, Veeraraghavan:2023vxd}. As a noble gas, radon can continuously emanate from the detector material and other components such as gas filters and pumps, and subsequently diffuse throughout the active target. Beta emitters from the $^{222}$Rn decay chain, such as $^{214}$Pb and $^{210}$Pb, can also contribute to a homogeneous background of low-energy events in the region of interest for rare event searches; while challenging to discriminate against \cite{XMASS:2017sid, Bunker:2020sxw, NEWS-G:2020fhm}, some data analysis approaches have shown promise in mitigating their impact \cite{XENON:2024lbh, LZ:2024zvo}. Although existing cryogenic methods---such as activated charcoal or copper traps \cite{ODwyer:2011, ABE201250, LZ:2020fty, Chen:2021hva, MicroBooNE:2022his, Li:2023lvs, Fatemighomi:2025gqd} and distillation  \cite{XENON100:2017gsw, Murra:2022mlr, Cui:2024ltd, XENON:2025nic}---have shown good radon removal performance, the increasing complexity and expansion to ton-scale experimental setups \cite{DarkSide-20k:2017zyg, Vahsen:2020pzb, Garcia-Viltres:2021swf, SNO:2021xpa, nEXO:2021ujk, JUNO:2021vlw, CUPID:2022wpt, KamLAND-Zen:2022tow, NEWS-G:2023qwh, Novella:2023gqb, Abe:2024kjq, McDonald:2024osu, GlobalArgonDarkMatter:2024wtv, XLZD:2024nsu, XLZD:2024pdv, PANDA-X:2024dlo, DUNE:2024wvj, Calgaro:2024boi, Bouet:2024neu} require the development of more efficient and practically optimized removal techniques.

We present a radon trap system developed for the New Experiments with Spheres-Gas (NEWS-G) dark matter experiment, currently achieving world-leading dark matter spin-dependent coupling to protons sensitivity for masses between 0.17 and 1.2 GeV/c$^2$ \cite{NEWS-G:2022kon, NEWS-G:2024jms}. The detector consists of a spherical proportional counter (SPC) filled with a noble gas mixture. It is equipped with a high-voltage (HV) sensor in the middle to achieve high avalanche amplification gain while having a linear energy response and low electronic noise \cite{Brossard:2019eby, Dastgheibi-Fard:2019nwy}. SPCs are capable of observing the energy response of single electrons \cite{NEWS-G:2019lqz}, which make them ideal detectors for searching for sub-GeV/$c^2$ dark matter candidates. However, electronegative impurities in the gas---such as water and oxygen---absorb primary electrons and compromise energy signal reconstruction through a process called attachment \cite{NEWS-G:2017pxg, Brossard:2023}. To address this, the NEWS-G gas handling system includes a gas purifier (or getter) to remove these impurities. Unfortunately, the getter also releases significant amounts of radon, with its daughter elements depositing on the inner surface  of the SPC and causing low-energy background events in the region of interest for dark matter searches \cite{NEWS-G:2022kon}. Traditional methods like active detector shielding and fiducialization cannot fully reject them. 

For our study, we evaluate the radon removal performance of silver-zeolite, specifically Ag-ETS-10~\cite{ExtraEdmonton1, ExtraEdmonton2}, compared to the commonly used coconut shell granular activated charcoal (Silcarbon Aktivkohle GmbH, Part \#: K835), both tested at room temperature. The rate of radon and its daughter emitters are monitored using a small-scale SPC detector filled with an argon/methane mixture, integrated within a closed-loop circulation system. We show that, at room temperature, the silver-zeolite significantly outperforms activated charcoal by three orders of magnitude in radon reduction, with near-complete removal of radon from the system.

Our work notably extends previous studies on radon trapping with silver-zeolite Ag-ETS-10. The efficacy of various silver-exchanged zeolites has been demonstrated in the important works \cite{Obrien:2021, Heinitz:2023, 10.1093/ptep/ptad160, Ogawa:2022tku, Sone:2024gox, 10.1093/ptep/ptaf007}, with silver-zeolite Ag-ETS-10 emerging as one of the most promising adsorbents for radon mitigation at room temperature \cite{ Heinitz:2023, 10.1093/ptep/ptad160}. These earlier studies on Ag-ETS-10 have been conducted with open-loop systems using gases such as nitrogen, air, or argon and employed a variety of detection methods, including the RadonEye detector and the alphaGUARD monitor. While those detectors are efficient for radon measurement, our work constitutes the first in-situ application of a silver-zeolite trap to validate its performance under realistic operating conditions using in a dark matter direct detection-type experiment with an SPC. In addition, near background-free measurements are obtained with our SPC, which also has high detection efficiency for radon decays due to the large active volume of the detector and low energy threshold. Moreover, our novel closed-loop experimental setup, in contrast to previous methods of Refs. \cite{ Heinitz:2023, 10.1093/ptep/ptad160}, shows how the performance of silver-zeolite Ag-ETS-10 traps for radon removal would be in dark matter detectors like NEWS-G and other rare event searches experiments, providing a distinct advantage over other configurations.

\section{Measurement campaigns}
\subsection{Experimental setup}
\begin{figure}
    \centering
    \includegraphics[width=0.8\textwidth]{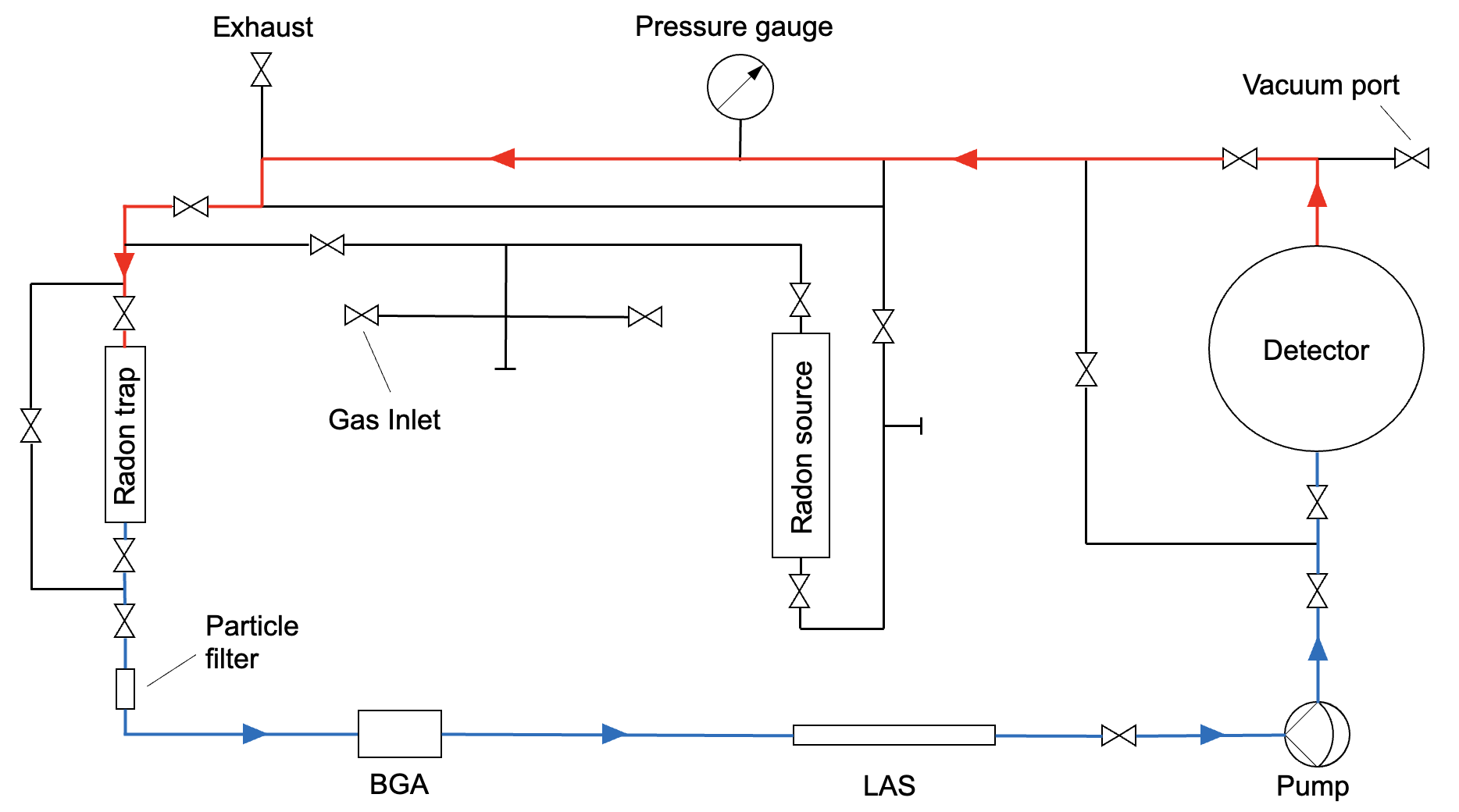}
    \caption{Closed-loop circulation system for the radon removal campaigns, made of stainless steel fittings, valves and tubing. The setup includes a radon source to diffuse radon into the SPC detector. The gas from the detector (red) circulates through the trap, where radon is adsorbed. The purified gas (blue) is pumped back to the detector, passing through a particle filter, a binary gas analyzer (BGA), and a custom-made laser absorption spectroscopy system (LAS) \cite{Obrien:2021, Garrah:2023}.}
    \label{fig:Experimental_setup}
\end{figure}

We monitor the event detection rate in a 30 cm diameter stainless steel SPC filled with 500 mbar of a 97\% argon and 3\% methane mixture. The detector is equipped with a two-mm diameter spherical anode at its center, biased to a positive HV of 1180 V via an insulated HV wire going through the grounded support rod. The operating conditions are selected to mostly contain the track lengths of alpha particles ($^{210}$Po (8.62 cm), $^{222}$Rn (9.09 cm), $^{218}$Po (10.3 cm), $^{214}$Po (15.1 cm), calculated with SRIM \cite{Ziegler:2010}) so that these deposit their full kinetic energy in the gas volume with high amplification gain. During each measurement, the SPC is integrated into a closed-loop circulation system consisting of a radon source, a radon trap, and a pump with a flow of 1 L/min, as shown in Fig.~\ref{fig:Experimental_setup}. The circulation ensures sustained radon capture by continuously passing the gas through the trap, while the purified gas is subsequently recirculated back into the detector.

The radon source (Pylon 1025) uses dry $^{226}$Ra to produce a calibrated quantity of $^{222}$Rn, with an equilibrium activity of 0.93 kBq and a continuous rate of 117.49 mBq/s, referring to the amount of $^{222}$Rn gas released from the source per unit of time. The radon trap consists of a custom 20 cm long stainless steel 1/2-inch tube, filled with 10 g of adsorbent material (silver-zeolite or activated charcoal) and capped at both ends with multi-purpose fiberglass. In addition, a particle filter (60 ${\upmu}$m pore size) is installed after the trap to prevent any granules of the adsorbent material from entering sensitive equipment within the circulation system. A binary gas analyzer (BGA) and a custom-made laser absorption spectroscopy system (LAS) are included to monitor the methane concentration in the gas during measurements \cite{Obrien:2021, Garrah:2023, Durnford:2024}. Because of the similar size of the radon (2.5\unit{\angstrom}--4.9\unit{\angstrom} \cite{Ezeribe:2017phs}) and methane molecules (3.8\unit{\angstrom}--4.1\unit{\angstrom} \cite{LI2025135593, methane}), the adsorbent also captures methane. This can alter the ionization properties of the gas, decreasing the gain in the SPC and affecting the amount of hydrogen target material sensitive to dark matter interactions. However, at room temperature, the amount of methane adsorbed (less than 20\%, as measured with the LAS) does not affect the results presented in this paper due to the high-energy nature of the alpha events.  

Before each measurement, the trap undergoes thermal regeneration at 160$^{\circ}$C within the closed-loop system.  This regeneration, also known as activation, removes any adsorbed gases and contaminants. It consists of flushing the trap with ultra-high-purity nitrogen gas for 20 minutes via the gas inlet to the exhaust, while heating it to 160$^{\circ}$C using a heating tape. The temperature is monitored with a thermocouple attached to the exterior surface of the trap. The nitrogen flow rate is controlled by adjusting the nitrogen supply pressure to 15--20 PSI using the regulator on the gas cylinder. After flushing, the trap is evacuated for another 20 minutes while maintaining the temperature at 160$^{\circ}$C. Once this process is completed, we stop the heating and maintain the vacuum in the trap until it cools to room temperature. Finally, the valves are closed to isolate the trap from the circulation system. This process is used for both silver-zeolite and activated charcoal. To further minimize contaminants in the gas loop, the entire system is evacuated to approximately $10^{-7}$ Torr, except the circulation pump, which cannot achieve a vacuum lower than $10^{-3}$ Torr. Finally, the gas mixture is injected, filling the system to a pressure of 500 mbar.

During the tests, the closed‑loop system remains sealed and leak‑tight, and the adsorbents (silver‑zeolite and activated charcoal) are never exposed to ambient air. The nitrogen used to regenerate the trap is ultra‑high‑purity grade 5 gas (N$_2$ > 99.999\%), with impurity levels of H$_2$O < 3 ppm, O$_2$ < 3 ppm and hydrocarbon < 0.1 ppm (Linde, Part \#: NI 5.0UH-T). The argon/methane mixture is a custom pre‑mix supplied by Linde; it is a non‑certified blend prepared from grade 5 argon and methane, with typically H$_2$O < 10 ppm.

\subsection{Experimental procedure}
We conduct three experimental campaigns to test the efficacy of the radon trap at room temperature and demonstrate reproducibility: two with silver-zeolite (campaigns 1 and 2) and one using activated charcoal (campaign 3). To ensure consistency across the campaigns, we organize the data collection into four phases: 

\begin{itemize}
  \item[$\bullet$] Background (phase I): initially, we perform at least a 24-hour background run to determine the minimal sensitivity of the detector; 
  \item[$\bullet$] Radon diffusion (phase II): radon is injected by diffusion exclusively into the detector for a day until reaching a raw event rate of $\sim$ 75 Hz. Once this rate is achieved, we close the radon source and isolate the detector. Continuing diffusion of radon from the gas system eventually raises the rate to $\gtrsim$ 100 Hz. During this period, the rest of the system remains closed and there is no circulation;
  \item[$\bullet$] Radon decay (phase III): we monitor the radon decay rate for approximately two days, while the rest of the system remains closed and there is no circulation;
  \item[$\bullet$] Trap open (phase IV): the closed-loop system, including the trap, is opened to the detector and the pump is used to circulate the gas, as shown in Fig.~\ref{fig:Experimental_setup}. Radon is removed at this step.
\end{itemize}

\begin{figure*}
    \centering
    \includegraphics[width=1\textwidth]{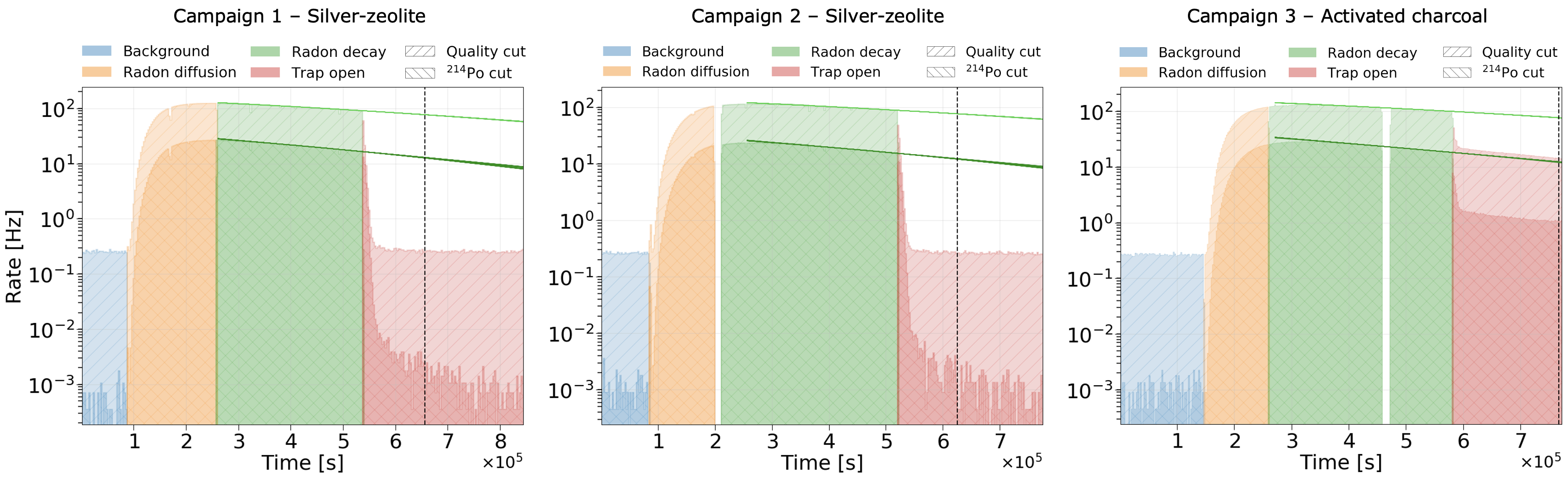}
    \caption{Measured rate with either quality cut only or with also the $^{214}$Po cut applied during the three radon campaigns, with the trap filled with either silver-zeolite or activated charcoal, at room temperature. The blue distribution corresponds to the background run (phase I), orange is the radon diffusion in the SPC (phase II), green is the radon decay (phase III), and red is the events recorded when the radon trap is open (phase IV). The green curves are the fitted/extrapolated decay rates for each campaign, depicting many random Markov Chain Monte Carlo (MCMC) samples. The light and dark shades make the distinction between quality cut only and with also the $^{214}$Po cut applied, respectively. The vertical dashed line defines the reference time where the expected rate with quality cut is equal to 76.5 Hz. Note that there are some visible gaps in the data (campaigns 2 and 3) due to brief pauses in data collection.}
    \label{fig:Campaign_rate_all}
\end{figure*}

The four phases for the three campaigns are shown in Fig.~\ref{fig:Campaign_rate_all}. The same procedure was used and the same trap was filled with 10 g of silver zeolite or activated charcoal adsorbent for each campaign.

\section{Data analysis methods}
To interpret the data and assess the experimental results, data selection cuts are applied, and the rate of radon events over time is determined. The performance of the trap is evaluated by comparing the expected rate from phase III without the intervention of the trap and the rate observed in phase IV.

\subsection{Quality cut}
\begin{figure}
    \centering
    \includegraphics[width=0.6\textwidth]{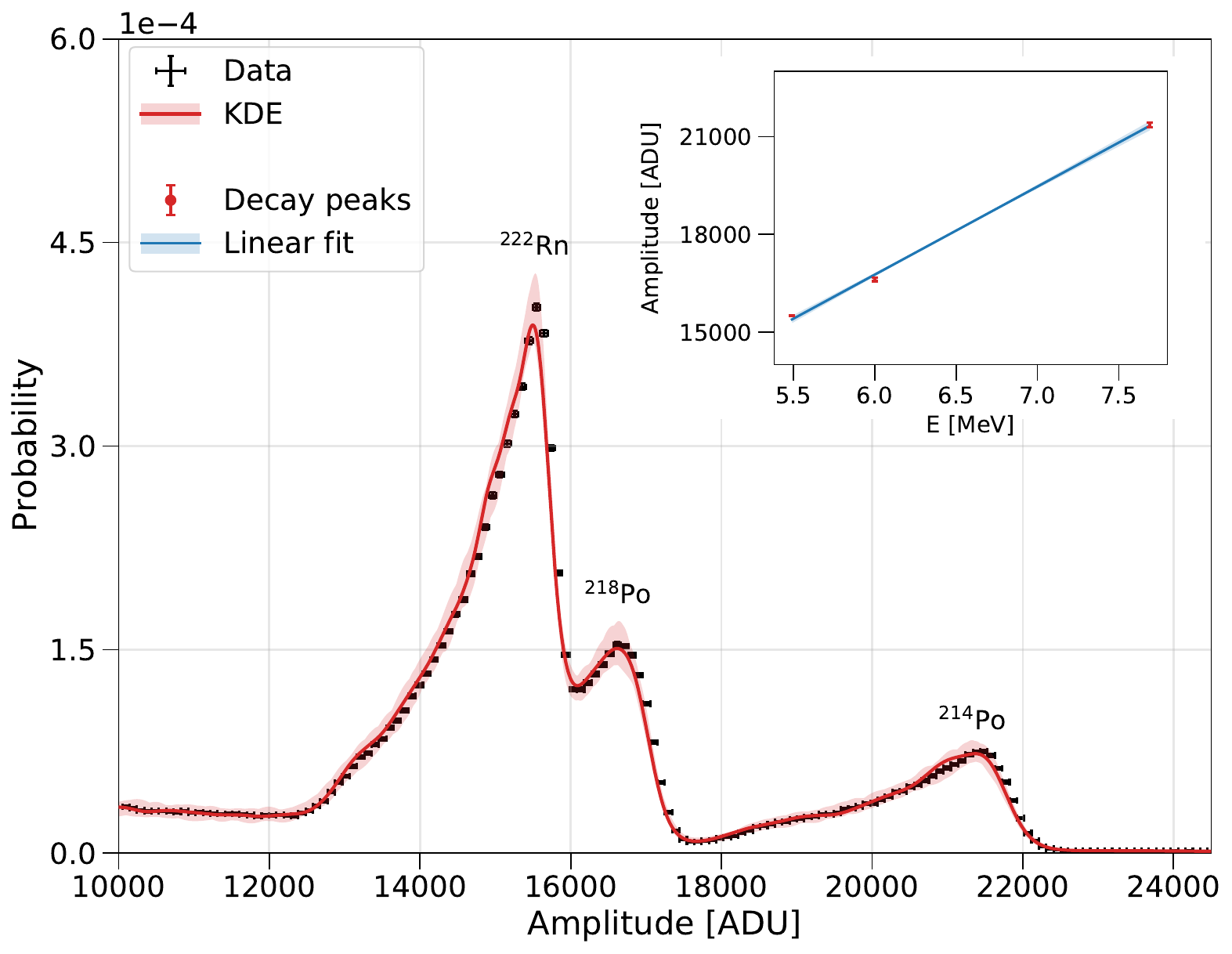}
    \caption{Amplitude distribution of events for approximately 6 hours of data in phase II of campaign 1, recorded by the SPC (black markers). The distribution is modeled with an adaptive-bandwidth Gaussian kernel density estimation \cite{silverman, wang2011bandwidth,kde_article,kde_code} with bootstrap-estimated \cite{efron} statistical uncertainty (red curve and $1\sigma$ shaded band). The inset shows the amplitude peak positions corresponding to the maximum energy deposition of three alpha decays versus the corresponding alpha energies (red markers with error bars), with a linear fit (blue curve and shaded band). A peak corresponding to $^{210}$Po might be present, but its rate is more than two orders of magnitude lower than the $^{222}$Rn peak with which it overlaps.}
    \label{fig:energy}
\end{figure}

We apply a quality cut that includes a loose pulse shape discrimination cut to remove non-physical events with characteristically short risetimes \cite{NEWS-G:2017pxg}, as well as requiring pulse amplitudes above 1200 ADU (arbitrary digital units). The latter removes most of the background events from atmospheric muons, as well as additional electronic noise events, selecting primarily the full energy alpha decays of $^{222}$Rn (5.5 MeV \cite{nndc_218}) and its daughter isotopes, $^{218}$Po (6.0 MeV \cite{nndc_214}) and $^{214}$Po (7.7 MeV \cite{nndc_210}), which have clear amplitude peaks above 12,000 ADU (see Fig.~\ref{fig:energy}). The rate over time in all three campaigns with this cut applied is shown in Fig. \ref{fig:Campaign_rate_all}. As these three alpha-decay processes quickly reach secular equilibrium, the proportion of events due to radon itself is approximately constant throughout each measurement campaign. Additionally, high energy beta decay events from the $^{222}$Rn decay chain ($^{214}$Bi in particular \cite{nndc_214}) are present in this data selection, but are also in equilibrium with the rest of the decay chain. 

Some degraded alpha events that do not deposit their full kinetic energy also fall below this quality cut. However, the fraction of affected alpha events remains approximately constant---and small---over time, increasing slightly as the detector gain decreases. A separate study quantifying the impact of this effect on signal efficiency due to this cut shows a drop of only $\sim$ 1\% over time (see Fig. 3.59 in Ref.~\cite{Durnford:2024}).

The constant background remaining in phase I for all campaigns (see Fig.~\ref{fig:Campaign_rate_all}) comes predominantly from the alpha decay of $^{210}$Po (5.3 MeV \cite{nndc_206}), originating from radioisotope depositions on the inner surface of the detector material. The SPC, made of stainless steel, does not have particular precautions to prevent $^{210}$Po contamination, but can nonetheless serve as a useful low-rate calibration source for this analysis. In phase IV, following the opening of the trap, the rate in campaigns 1 and 2 decreases significantly over several hours, eventually reaching the $^{210}$Po alpha background level observed in phase I. For campaign 3, using activated charcoal, the rate remains up to two orders of magnitude higher compared to the rate in phase I and exhibits exponential decay, indicating the continued presence of radon.

\subsection{$^{214}$Po cut}

To achieve a lower background rate than what remains after applying the quality cut, an additional stricter selection is applied to select only $^{214}$Po events. $^{214}$Po decays achieve secular equilibrium with $^{222}$Rn due to its short half-life (164 ${\upmu}$s \cite{nndc_210}) and the short half-lives of the other intermediate isotopes compared to $^{222}$Rn and its subsequent daughter isotopes.  As shown in Fig. \ref{fig:Campaign_rate_all}, the selection of only $^{214}$Po events significantly improves our signal-to-background ratio and allows us to assess our results with near-zero background in later analyses, as there is no $^{210}$Po background in this amplitude regime.

\begin{figure}[h!]
    \centering
    \includegraphics[width=0.7\textwidth]{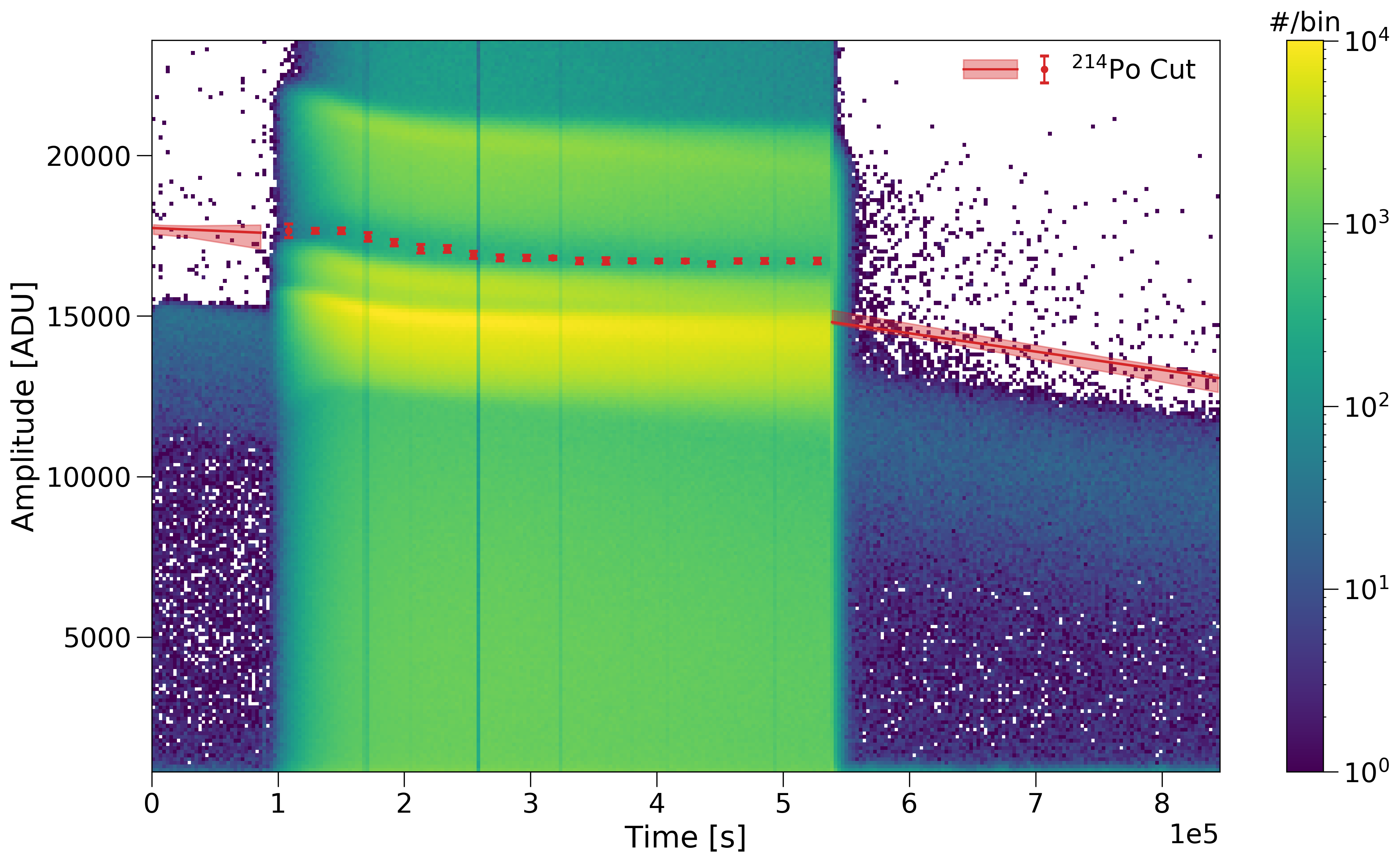}
    \caption{2D histogram of the amplitude distribution over time in campaign 1 with the trend of $^{214}$Po cut  obtained across all phases. In phases II and III, the $^{214}$Po cut is defined as the amplitude minima between the $^{218}$Po and $^{214}$Po peaks in time bins of approximately 6 hours duration, indicated as red markers with statistical error bars. In phases I and IV, a linear fit of the $^{210}$Po peak position over time is scaled to match the $^{214}$Po cut at the beginning of phase II/end of phase III, shown as red curves with shaded $1\sigma$ uncertainties.}
    \label{fig:Po214cut}
\end{figure}

When radon is injected in phases II and III, the amplitude distribution exhibits different peaks corresponding to $^{222}$Rn (5.5 MeV \cite{nndc_218}) and the daughter isotopes $^{218}$Po (6.0 MeV \cite{nndc_214}) and $^{214}$Po (7.7 MeV \cite{nndc_210}), proportional to the energy of each alpha emitter (see Fig.~\ref{fig:energy}). The $^{214}$Po cut applied is identified by the time-evolving minima between the $^{218}$Po and $^{214}$Po peaks, within approximately 6-hour data segments. This method cannot be applied directly to phase I of all campaigns and phase IV of campaigns 1 and 2, since there is no $^{214}$Po amplitude peak feature. Instead, during these phases, the $^{210}$Po amplitude peak is identified within similar 6-hour data segments, fit with a linear function over time, and scaled to match the amplitude corresponding to the $^{214}$Po cut defined in phases II and III, assuming energy linearity in this regime. An example of the evolution of the $^{214}$Po amplitude cut across all phases for campaign 1 is shown in Fig.~\ref{fig:Po214cut}.

To verify the linear energy response of the SPC in this regime assumed to perform the scaling described above, amplitude spectra---such as the one shown in Fig.~\ref{fig:energy}---are characterized with an adaptive-bandwidth Gaussian kernel density estimation \cite{silverman, wang2011bandwidth,kde_article,kde_code}, with statistical uncertainties being estimated using a bootstrapping procedure \cite{efron}. The resulting  peak amplitudes are well-fitted by a linear function of the corresponding alpha decay energies, as shown in the inset of Fig.~\ref{fig:energy}.

Due to the large amount of radon remaining in phase IV of campaign 3 (see Fig.~\ref{fig:Campaign_rate_all}), the $^{214}$Po amplitude cut is determined using a similar method as in phases II and III. To obtain a time-dependent $^{214}$Po amplitude cut for phase IV, the amplitude minima between the $^{218}$Po and $^{214}$Po peaks are found in approximately 6-hour data segments. The bootstrap resampling \cite{efron} is then applied to estimate the statistical uncertainty for each point. These points are then fit with a linear function to define the $^{214}$Po amplitude cut for phase IV.

\subsection{Expected decay rates}
\label{ss:expected}

Evaluating the performance of the adsorbent requires determining the expected decay rates of radon and its daughters over time without the intervention of the trap. The expected decay rates with either just the quality cut or also with the $^{214}$Po cut applied are determined using the phase III data of each campaign. However, operating the SPC at event rates near 100 Hz with a 2 ms event acquisition time window introduces a significant probability of coincidental events, or pileup, and causes dead-time losses due to limitations in the data acquisition (DAQ) system \cite{Durnford:2024}. These effects reduce the observed decay rate compared to the true rate and introduce potential sources of systematic error in subsequent analyzes. It is therefore necessary to infer the true radon rate from the measured rate, accounting for pileup and dead-time losses.

To achieve this, a Monte Carlo (MC) simulation of radon decay is produced for a given radon rate in addition to a given background rate, which includes the expected pileup and dead-time effects to reflect the actual conditions of our experiment. Any simulated decays occurring within 1 ms of each other (half the duration of a single event window) are classified as coincident events (resulting in a single recorded event). Simulated decays occurring within 1 to 2 ms of a preceding pulse are considered lost due to DAQ deadtime. This treatment accounts for the DAQ software centering the maximum amplitude of the triggered event trace within its acquisition time window, ensuring that no two pulses can be more than half an event window apart and still be contained in the event window. High-statistics MCs are produced for true radon rates ranging from approximately 30 Hz to 200 Hz, along with the measured $^{210}$Po background rate (phase I) for each campaign. Each MC dataset spans a duration longer than the real data from phase III, covering the time-spans of both phase III and phase IV. The resulting log-scale rate trends---separated into single decay event vs.\ pileup events---are characterized using univariate splines \cite{scipy_spline}.

The results of the decay rate MCs are incorporated into a joint model for the observed event rate over time, with and without the application of the $^{214}$Po cut. The free parameters of the model define: 
\begin{itemize}
    \item[(1)] the true radon decay rate at the start of each phase III fit, from the discrete sets of values for which MCs are produced;
    \item[(2)] the overall alpha event signal efficiency with the quality cut applied ($f_{\alpha}$);
    \item[(3)] signal efficiency of observed $^{214}$Po events with the $^{214}$Po cut applied ($f_{\text{Po-214}}$); 
    \item[(4)] the fraction of pileup events that pass the quality or $^{214}$Po cut ($f_{\text{pileup}}$ and $f_{\text{pileup, Po-214}}$, respectively).
\end{itemize}   Adding the background rates for each data selection ($\eta$ and $\eta_{\text{Po-214}}$), the expected event rates over time $\nu$, with  and without the  $^{214}$Po cut, are given as
\begin{align}
\nu&(t) = f_{\alpha} \left[ 10^{y_{\text{Rn-222}}(t)} + 10^{y_{\text{Po-218}}(t)} + 10^{y_{\text{Po-214}}(t)} \right]  + f_{\text{pileup}} \times 10^{y_{\text{pileup}}(t)} + \eta, 
\nonumber {\color{white} \frac{|}{|}} \\
\nu&_{\text{Po-214}}(t) = f_{\text{Po-214}} \times 10^{y_{\text{Po-214}}(t)} + f_{\text{pileup,\,Po-214}} \times 10^{y_{\text{pileup}}(t)} + \eta_{\text{Po-214}},
\label{eq:phaseIII_nu}
\end{align}
\noindent where $y_{\text{Rn-222}}(t)$, $y_{\text{Po-218}}(t)$, $y_{\text{Po-214}}(t)$, and $y_{\text{pileup}}(t)$ are the interpolated values from the decay MC splines for the corresponding radon decay chain isotopes and pileup events, respectively.

A joint, binned, Poisson log-likelihood function is derived for the event rate over time given this model;
\begin{align}
\log &\mathcal{L}\ = \sum_{i}\left[ h_{3}(i)\   \log\nu(t_i) - \nu(t_i)\right] + \sum_{i}\left[ h_{3}^{\text{Po-214}}(i)\ \log\nu_{\text{Po-214}}(t_i) - \nu_{\text{Po-214}}(t_i)\right],
\end{align}
\noindent where $h_3(i)$ and $h_{3}^{\text{Po-214}}(i)$ are the observed event counts in phase III in time bin $i$ (with central time $t_i$) with and without the $^{214}$Po cut applied, respectively. This model is fit to the data using a Markov Chain Monte Carlo (MCMC). Specifically, the \python{} package \verb|emcee| is used with the ``stretch-move'' Metropolis-Hastings algorithm \cite{GW10,emcee}, propagating the 100 random walkers for $2\times10^5$ steps after an initial burn-in phase of $4\times10^4$ steps. This exceeds the recommendation that the number be at least $50 \times$ the estimated autocorrelation time \cite{emcee}. 
The resulting fit of the observed event rates with each cut applied is shown in Fig.~\ref{fig:Campaign_rate_all}, including extrapolations to phase IV for each campaign to show the expected decay rates without the intervention of the radon trap.

\section{Results and discussion}

\subsection{Rate comparison}

Figure \ref{fig:Campaign_rate_all} clearly demonstrates that campaigns 1 and 2, which utilize the silver-zeolite trap, outperform campaign 3 using activated charcoal in terms of radon and $^{214}$Po event reduction. Indeed, the rates in phase IV drop significantly before stabilizing. In contrast, the drop in rate is much smaller in phase IV of campaign 3, with an exponential decay indicating the presence of radon remaining in the detector.

\begin{figure}
    \centering
    \includegraphics[width=0.6\textwidth]{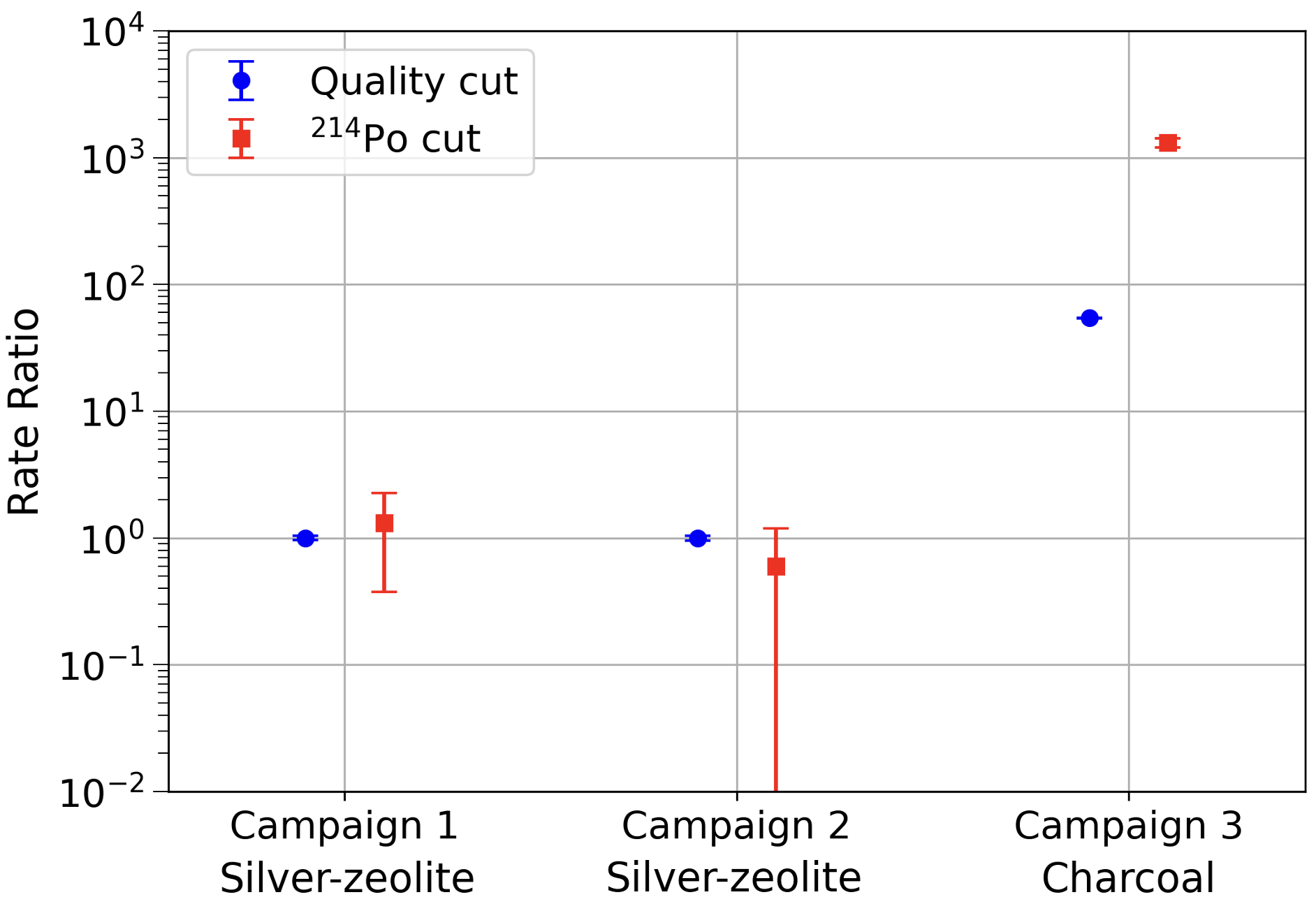}
    \caption{Ratio between the rate in phase IV (trap open) at the reference time and the rate in phase I (background run) for each campaign with the trap filled with silver-zeolite (campaigns 1 and 2) and activated charcoal (campaign 3). The blue circle and red square represent the rate ratio obtained with either just the quality cut or with $^{214}$Po cut applied, respectively. Complete radon removal correspond to a rate ratio of 1.}
    \label{fig:Rate_ratio}
\end{figure}

As an initial assessment of the trap performance, we compare the rates obtained in phases I and IV with and without the $^{214}$Po cut throughout the three campaigns. As can be seen in Fig.~\ref{fig:Campaign_rate_all} the rate in phase I for all campaigns is approximately constant, and is fitted with a constant function. To compare this rate with the rate in phase IV after the trap is opened, we define the following reference time. We extrapolate the expected rate from phase III (without trap) with quality cut to later times and determine the reference time where the extrapolation reaches 76.5 Hz (dashed vertical lines in Fig.~\ref{fig:Campaign_rate_all}). Then, the constant rate in phase IV of campaigns 1 and 2 is determined by the average count of all time bins starting from the reference time. However, in campaign 3, the rate decreases exponentially and cannot be fitted with a constant function. To facilitate comparison with campaigns 1 and 2, we also determine the rate in phase IV at the reference time for all campaigns. Table~\ref{tab:rate_all} summarizes the rate in phases I and IV with the quality and $^{214}$Po cuts, as well as the rate in phase IV for all campaigns, calculated at the reference time. From this, we calculate the rate ratio between phase IV and phase I with either the quality or $^{214}$Po cut, as shown in Fig.~\ref{fig:Rate_ratio}.

\begin{table}
\centering
  \begin{threeparttable}
  \caption{Rate in phase I (background) and phase IV (trap open) with either the quality or $^{214}$Po cut for all campaigns. The rate in phase IV obtained from the constant fit and at the reference time are also included for all campaigns.}
  \label{tab:rate_all}
  \centering
\begin{tabularx}{\textwidth}{w{c}{2.5cm}w{c}{3.7cm}w{c}{3.7cm}w{c}{3.7cm}}

\hline\hline

& & \\ [-2.7ex]
Rate (Hz) & Phase I  & Phase IV \tnote{a} & Phase IV  \tnote{b} \\ [0.2ex]

\hline\hline
& & \\ [-2ex]
\textbf{Quality Cut} & & \\ [0.5ex]
Campaign 1  & 0.252 $\pm$ 0.002 & 0.254 $\pm$ 0.001  & 0.251 $\pm$ 0.009  \\ [0.5ex]
\hline
& & \\ [-2ex]
Campaign 2  & 0.262 $\pm$ 0.002 & 0.260 $\pm$ 0.001    & 0.262 $\pm$ 0.011  \\  [0.5ex]
\hline
& & \\ [-2ex]
Campaign 3   & 0.266 $\pm$ 0.001 &  -- & 14.5 $\pm$ 0.08 \\[0.5ex]
\hline\hline
& & \\ [-2ex]
\textbf{$^{214}$Po cut} & & \\ [0.5ex]
Campaign 1  & (5.36 $\pm$ 0.62)$\times$10$^{-4}$ & (7.58 $\pm$ 0.66)$\times$10$^{-4}$  & (7.07 $\pm$ 5.00)$\times$10$^{-4}$  \\  [0.5ex]
\hline
& & \\ [-2ex]
Campaign 2  & (7.55 $\pm$ 0.78)$\times$10$^{-4}$ & (8.63 $\pm$ 0.86)$\times$10$^{-4}$  & (4.50 $\pm$ 4.50)$\times$10$^{-4}$  \\  [0.5ex]
\hline
& & \\ [-2ex]
Campaign 3  & (8.18 $\pm$ 0.67)$\times$10$^{-4}$ & --  &  1.07 $\pm$ 0.02 \\[0.5ex]

\hline\hline
\end{tabularx}

 \begin{tablenotes}
       \item [a] Rate calculated from constant fit after the reference time.
       \item [b] Rate calculated at the reference time.
     \end{tablenotes}
  \end{threeparttable}
  
\end{table}

Campaigns 1 and 2, using the silver-zeolite trap, consistently achieve near complete radon removal, whereas the radon levels in the activated charcoal campaign 3 are two to three orders of magnitude higher. As expected, the rate ratio obtained with the $^{214}$Po cut is typically larger, but more affected by statistical fluctuations due to the small rate of events in phase I and phase IV.

From Table \ref{tab:rate_all}, the rates in phase I with the quality cut are nearly consistent, with campaigns 2 and 3 each exhibiting a slightly higher rate than the previous campaign. This small increase may be attributed to the buildup of $^{210}$Po on the inner surface or volume of the SPC due to consecutive injections of radon over time. Based on the expected buildup and the lifetime of $^{210}$Po, a small increase in rate—on the order of 0.01 Hz—is expected, which aligns well with the observed data. Similarly, for the $^{214}$Po cut, the rates align closely within the uncertainties with a slight increase, likely due to the presence of additional $^{210}$Po, contributing to more mis-reconstructed events. The same trend is observed in phase IV when comparing the rates of campaigns 1 and 2 with the quality and $^{214}$Po cuts. Nonetheless, the ratio between the rates in phase IV and phase I for campaigns 1 and 2 approaches 1 (see Fig.~\ref{fig:Rate_ratio}), indicating that the radon rate quickly reduces to pre-injection levels in campaigns 1 and 2. This also demonstrates that the performance of the radon trap aligns well with the achievable sensitivity of our experimental setup.

\subsection{Radon reduction ratio calculation}

To further quantify the performance of the trap, we calculate the radon reduction ratio $R$ (or R-value) from phase IV, defined as
\begin{align}
            R (t) = \frac{R_{\text{exp}}(t)} { R_{\text{obs}}(t)}.
\label{eq:Radon_Reduction}
\end{align}
\noindent Here, $R_{\text{exp}}(t)$ is the expected rate obtained from the extrapolated rate trend (see Sec.~\ref{ss:expected}) and $R_{\text{obs}}(t)$ is the observed rate from the experimental data after the trap is opened in phase IV. The latter is fit with a generic model consisting of an exponential function plus a constant component, in order to smooth--over statistical fluctuations and remove dependence on the binning of the counts over time. This fit was done with an MCMC following the same procedure to obtain the expected rate trend described in Sec. \ref{ss:expected}. 

The R-value at a given time in phase IV is determined by optimizing a Poisson likelihood (with a Nelder-Mead algorithm~\cite{Virtanen:2019joe, Gao:2012guu}) comparing the observed count model with the expected count rate scaled by $R(t)^{-1}$. Specifically, separate results with and without the $^{214}$Po cut are obtained with a log-likelihood function for the phase IV model of each campaign, at the central time of each bin $i$, scaling the phase III decay rate trend by $R(t_i)^{-1}$ and allowing the background rate $\eta$ to vary as a nuisance parameter, such as
\begin{align}
\log \mathcal{L}_R(i)\ =\ & h_4(i) \log\nu(t_i) - \nu(t_i)  - \frac{1}{2}\ \left(\frac{\eta-\eta_{\mathrm{obs}}}{\sigma_{\eta}}\right)^2.
\label{eq:L1}
\end{align}
\noindent The latter term is a Gaussian constraint term for the background rate, with respect to the observed background rate (from phase I) and its uncertainty ($\eta_{\mathrm{obs}}$ and $\sigma_{\eta}$, respectively). $h_4(i)$ is the observed count rate model for phase IV in time bin $i$. The expected counts in each time bin, $\nu(t_i)$, is derived from the expected rate trend from phase III (see Eq.~\ref{eq:phaseIII_nu}) multiplied by $R(t_i)^{-1}$, plus $\eta$. Given the expected 15.1 cm average track length of $^{214}$Po events \cite{Ziegler:2010} relative to the 30 cm diameter of the SPC, some events may not deposit all their energy within the detector volume. Further, some fraction of $^{214}$Po isotopes are expected to settle on the inner surface of the detector before decaying, increasing the fraction of events that will not deposit their full energy in the SPC gas. This could result in an underestimation of the actual rate of $^{214}$Po. Consequently, the expected and observed $^{214}$Po event rates in phase IV are both conservatively reduced by 50\% in the calculation of $R(t_i)^{-1}$, increasing the statistical uncertainty in this result. 

\begin{figure*}
    \centering
    \includegraphics[width=1\textwidth]{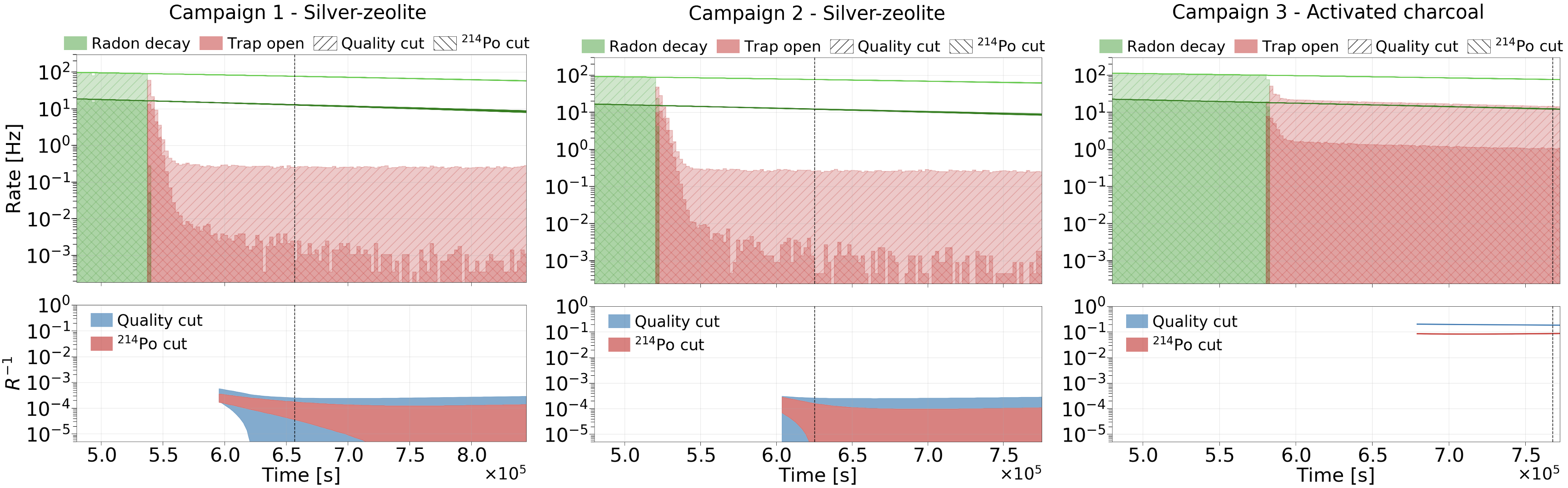}
    \caption{\textit{Top panel} Expected decay rates with either the quality cut (light green curves) or with the addition of the $^{214}$Po cut applied (dark green curves), derived from modeling of the phase III data from each campaign (green histograms), as well as the phase IV data (red histograms). \textit{Bottom panel} The 90\% confidence level (CL) intervals of the inverse R-values over time, obtained with the quality cut (blue bands) and $^{214}$Po cut (50\% reduced signal efficiency) additionally applied (red bands). The vertical dashed line defines the reference time where the extrapolated rate with quality cut (light green line) is equal to 76.5 Hz.}
    \label{fig:All_RValues}
\end{figure*}

\begin{figure}
    \centering
    \includegraphics[width=0.6\textwidth]{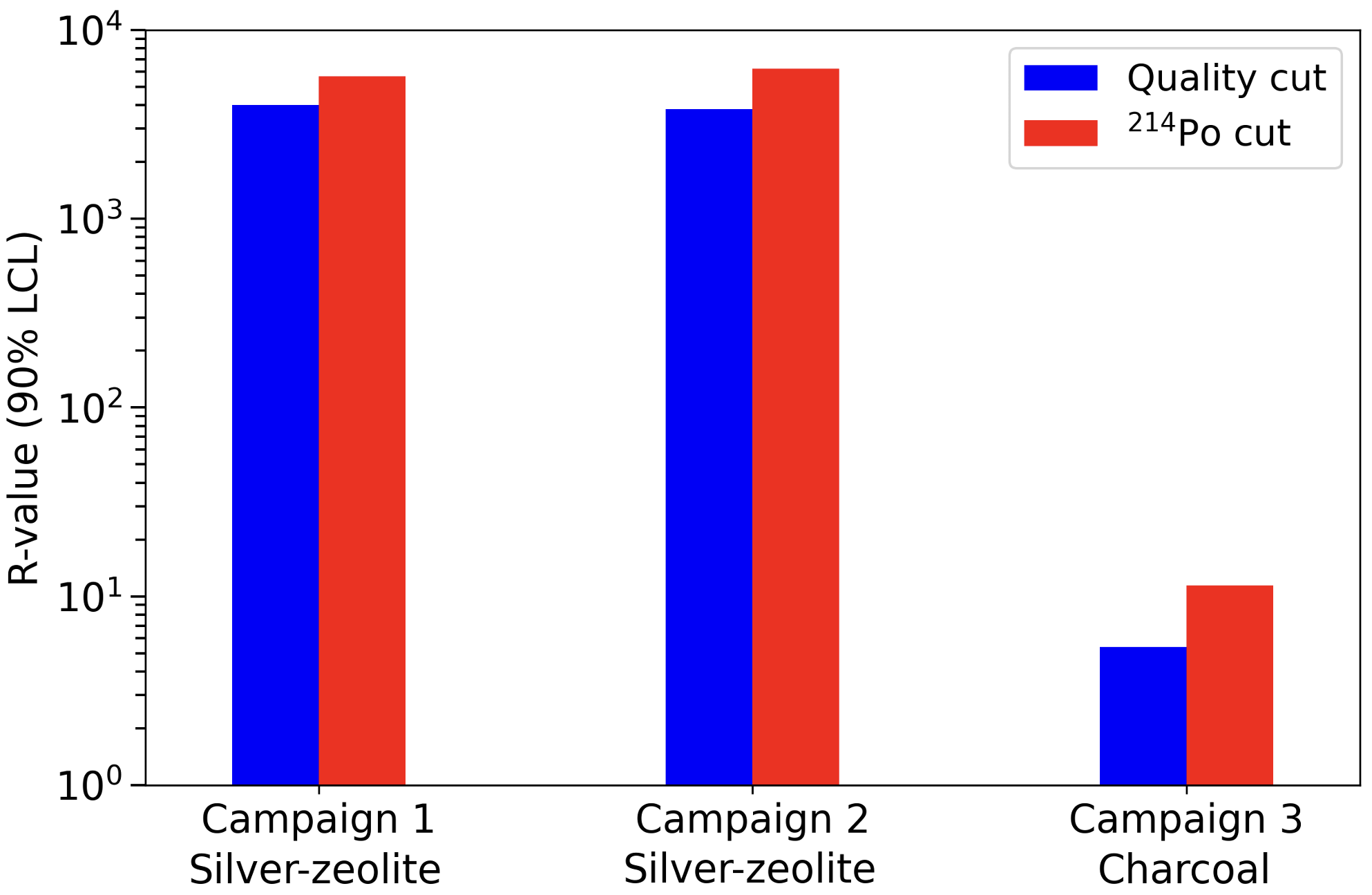}
    \caption{R-values at 90\% lower confidence limit (LCL), obtained at the reference time, with the trap filled with silver-zeolite (campaigns 1 and 2) and activated charcoal (campaign 3). The blue and red bars represent the R-values obtained with either just the quality cut or also with the $^{214}$Po cut applied (with 50\% reduced signal efficiency), respectively.}
    \label{fig:Rvalue_results}
\end{figure}

A $90\%$ confidence level (CL) interval is calculated for each time bin using the Feldman-Cousins method \cite{Feldman:1997qc}. This process is repeated for 500 randomly--drawn MCMC samples from the fit of the phase III rate model and the phase IV observed rate model to include statistical uncertainty for both fits. This yields an ensemble of slightly--varying 90\% CL intervals for $R(t_i)^{-1}$ for each time bin, of which we take the median result for each. The $R(t_i)^{-1}$ 90\% CL intervals are shown in the bottom panels of Fig.~\ref{fig:All_RValues} for each campaign. From these results, we determine the final R-value, $R(t)$, at the reference time, which is denoted by vertical dashed lines in Fig.~\ref{fig:All_RValues}. The derived values and their associated 90\% CL intervals (upper and lower) for both cuts are summarized in Table~\ref{tab:rvalue}. 
To quantitatively compare the R-value across all campaigns, the 90\% lower confidence limit (LCL) of $R$ at the reference time is reported as the final result for each campaign and shown in Fig.~\ref{fig:Rvalue_results} with or without the $^{214}$Po cut (with 50\% reduced signal efficiency) applied.

In campaigns utilizing silver-zeolite (campaigns 1 and 2), the 90\% LCL R-values range from 3.8$\times$10$^{3}$ to 6.2$\times$10$^{3}$, which is three orders of magnitude higher than those found in campaign 3, using activated charcoal, where R-values range from 5.4 to 11.4. Moreover, campaigns 1 and 2 demonstrate similar performance of silver-zeolite, as evidenced by their similar results. As anticipated, R-values derived with the $^{214}$Po cut applied are generally higher than those with the quality cut, due to a higher signal-to-background ratio.

\begin{table}[h!]
\centering
  \begin{threeparttable}
\centering
\caption{90\% CL intervals (lower and upper limits) of R-values at the reference time with only the quality cut or also with the $^{214}$Po cut (reduced by 50\%) applied for each campaign. The 90\% lower confidence limit (LCL) is reported as the final result for each campaign and shown in Fig.~\ref{fig:Rvalue_results}} 
\label{tab:rvalue}

\begin{tabularx}{1\textwidth}{w{c}{3.1cm}|w{c}{3.3cm}|w{c}{4.cm}|w{c}{3.3cm}}

\hline\hline

& & \\ [-2.5ex]
& Campaign 1 & Campaign 2 & Campaign 3 \\[0.5ex]

\hline
& & \\ [-2ex]
Quality cut & [4006.9, inf] & [3803.3, inf] & [5.4086, 5.5079] \\[1ex]

$^{214}$Po cut (50\%)  & [5684.9, 28337] & [6232.3, 2.0314$\times$10$^{6}$] & [11.395, 11.886] \\[0.5ex]

\hline\hline
\end{tabularx}
\end{threeparttable}

\end{table}

\subsection{K-factor estimation}

To enable a better comparison with previous studies \cite{10.1093/ptep/ptad160, Heinitz:2023}, we also estimate the adsorption constant, or \textit{K-factor} [m$^3$/kg] obtained with our gas mixture 97\% argon and 3\% methane. With our current setup and available data, we cannot apply the standard definitions used in Ref. \cite{VESELSKA2026134640}. Instead, we follow the approach of Ref. \cite{Sone:2024gox}, assuming that the radon concentration in our system---dominated by the detector volume---is proportional to the circulation time. In this framework, the R‑value is defined as the inverse of the radon concentration ratio (Eqs. 2--3 in \cite{Sone:2024gox}), which allows us to determine the retention time required to compute the estimation of the \textit{K-factor} for our gas mixture. 

With a detector volume of 14 L, a flow rate of 1 L/min, 10 g of adsorbent, and the 90\% LCL R‑values with quality cuts from Table \ref{tab:rvalue}, we obtain \textit{K} = 6.1720 m$^3$/kg for activated charcoal (campaign 3), and \textit{K} = 5608.3 and 5323.2 m$^3$/kg for silver-zeolite Ag-ETS-10 from campaigns 1 and 2, respectively.

The \textit{K-factor} obtained for activated charcoal is consistent with standard values at room temperature reported in the literature, typically in the range of \textit{K} = 3--10 m$^3$/kg \cite{VESELSKA2026134640}.  For the Ag-ETS-10, a result of \textit{K} = 5323.2--5608.3 m$^3$/kg is  comparable to those at room temperature reported in Ref. \cite{10.1093/ptep/ptad160} (\textit{K} = 1400 m$^3$/kg) and Ref. \cite{Heinitz:2023} (\textit{K} = 3400 m$^3$/kg) in nitrogen gas and (\textit{K} = 4300 m$^3$/kg) in air \cite{Heinitz:2023}. However, it remains significantly lower than the value measured in pure argon  (\textit{K} $\approx$ 20 000 m$^3$/kg) \cite{Heinitz:2023}. 

It is important to note that differences in detector technology and operating pressure prevent a direct comparison. In addition, the \textit{K-factor} strongly depends on the type of carrier gas, and no compatible measurements exist for our specific gas mixture or for the low operating pressure used here. Furthermore, the absence of detectable radon re‑emission in campaigns 1 and 2 with silver-zeolite (as observed in Ref. \cite{Heinitz:2023} for pure argon) suggests that the actual retention time may be significantly longer than our estimates. This would imply that our derived \textit{K-factor} is likely conservative. Because our result is an estimation rather than a fully validated measurement, we intend to adapt our closed-loop system for future campaigns---particularly those planned at lower temperatures---to measure the radon activity directly inside the trap and to extend the post-trap-opening measurement period. These improvements will provide a more consistent method for determining the \textit{K-factor} to compare with existing results.

\section{Conclusion and outlook}
A comparison of the silver-zeolite Ag-ETS-10 and activated charcoal campaigns reveals several critical insights. First, the silver-zeolite trap demonstrates significantly higher efficiency in radon removal, with R-values three orders of magnitude greater than those observed with the activated charcoal trap. This conclusion is compatible with the relative \textit{K‑factor} estimations derived from alternative method \cite{Sone:2024gox}, which reflect a similar contrast in performance between the two adsorbents. Additionally, the rate ratio, which compares event rates after the trap is opened to those before radon diffusion, approaches 1 for the silver-zeolite campaigns, demonstrating near-complete radon removal from the system. In contrast, the activated charcoal campaign exhibits residual radon levels in the rate ratio that are two to three orders of magnitude higher. Finally, in all measurements, the silver-zeolite campaigns show consistent performance, as evidenced by similar results across both campaigns, confirming reliable and reproducible outcomes.

We demonstrate that silver-zeolite Ag-ETS-10 achieves superior radon removal efficiency at room temperature compared to activated charcoal, which is typically used at cryogenic temperatures. Our innovative closed-loop system offers a novel approach beyond previous methods and is the first to demonstrate validated performance under realistic operating conditions, such as those present in dark-matter-type detectors. These findings not only confirm and extend previous studies \cite{Obrien:2021, Heinitz:2023, 10.1093/ptep/ptad160, Ogawa:2022tku, Sone:2024gox, 10.1093/ptep/ptaf007}, but also highlight the strong potential of silver-zeolite for ambient-temperature radon reduction systems in rare-event search experiments. This advancement is particularly important for underground laboratories, where elevated radon levels compromise the ultra-low background environments required for dark matter and neutrino physics experiments \cite{Cui:2023fcd, Huang:2024}. Integrating silver-zeolite filters into ventilation systems and clean rooms can substantially reduce radon concentrations and enhance the supporting capabilities of these facilities \cite{Kamaha:2022ooe, Baudis:2022qjb}. Moreover, eliminating the need for cryogenic cooling offers significant operational advantages by simplifying system design and reducing the overall complexity of radon reduction systems for future ton-scale detectors and beyond. The development of room-temperature radon filters could enable useful applications across multiple sectors, including residential basements, healthcare facilities, and other specialized industrial environments. Since our experimental campaigns with silver-zeolite Ag-ETS-10 have been conducted with an actual dark matter-type detector, they serve as a proof of principle for its application in current and future rare event searches.

\section*{Acknowledgment}

\noindent This work was supported by the Canada First Research Excellence Fund through the Arthur B. McDonald Canadian Astroparticle Physics Research Institute and through the University of Alberta startup fund UOFAB Startup Piro. M.-C.~Piro. also acknowledges funding from the Natural Sciences and Engineering Research Council of Canada (NSERC) Subatomic Physics Discovery Grant (project) No. SAPPJ-2024-00025, from which P.~Gros. and Y.~Deng. were further supported. The work of D.~Durnford. and C.~Garrah. was also supported by the NSERC Graduate Scholarships Doctoral and Master programs, respectively. We are grateful for the support of Compute Ontario \cite{ontario}, the BC DRI Group, and the Digital Research Alliance of Canada \cite{alliance} required to carry out the simulation needed for this manuscript. We thank Hervé Le Provost (IRFU-CEA/Saclay) for his software development work in our data acquisition system. We also thank the NEWS-G collaboration for their technical support provided for this work and the review of this manuscript. We also thank Daniel Kuznicki for assisting with the silver-zeolite Ag-ETS-10 sample obtained from Extraordinary Adsorbents Inc. of Edmonton, Alberta~\cite{ExtraEdmonton1}.

\bibliographystyle{ptephy}
\bibliography{PTEP_main}

\end{document}